\documentstyle[11pt,aaspp4]{article}
\begin{document} 
\title{Orbits in a Neighboring Dwarf Galaxy According to MOND} 
\author{Daniel M\"uller\altaffilmark{1} and 
Reuven Opher\altaffilmark{2}} 
\affil{Instituto Astron\^omico e Geof\'\i sico, Universidade de S\~ao Paulo\\ 
Av. Miguel St\'efano 4200, 04301-904 S\~ao Paulo, SP, Brazil} 
\altaffiltext{1}{muller@orion.iagusp.usp.br} 
\altaffiltext{2}{opher@orion.iagusp.usp.br} 
\begin{abstract} 
We study the orbits in the MOND theory within 
a dwarf galaxy of mass $M_d\sim 10^8M_\odot$ at a distance of $\sim 100$kpc 
from a neighboring galaxy of mass 
$M_g=5\times 10^{11} M_\odot$, such as ours. 
It is assumed that a second mass $m<<M_d$ is gravitationally bound to $M_d$   
by a previously calculated potential for the MOND theory. 
This potential is obtained for a free falling mass $M_d$ in 
a constant external gravitational 
acceleration field $\nabla\phi_g$. The numerical 
technique of surfaces of section is used to study the stability of the 
phase-space orbits in the dwarf galaxy. Equatorial orbits with sufficiently 
small eccentricities $e<0.65$ are found to be stable with respect to small changes in the 
initial conditions. (The equatorial plane is perpendicular to 
the direction of $\nabla\phi_g$, which is along the line joining $M_d$ and $M_g$.) 
For decreasing values of the conserved component of the angular momentum, in 
the direction of $\nabla\phi_g$, equatorial stability is lost. 
  
\end{abstract} 
\section{Introduction} 

The inconsistency between luminous and dynamical mass measurements 
is well known. An alternative way to explain this inconsistency 
without dark matter, is by Modified Nonrelativistic Dynamics (MOND) for  
galaxy and galactic systems \cite{M1} \cite{M2}, \cite{M3}. This theory was 
later 
developed into an alternative theory of gravitation by \cite{BM} (1984), 
which we hereafter  refer to as
the Bekenstein-Milgrom (BM) theory. MOND agrees with 
the \cite{TF} and \cite{FJ} laws, and explains the 
dynamics in elliptical and spiral galaxies, without the need for dark matter. 
It satisfies conservation of energy, momentum, angular momentum, as well as the 
weak equivalence principle (\cite{BM} 1984).

Recent studies of rotation disks of Low Surface 
Brightness (LSB) Galaxies by \cite{SG}, \cite{Block} and \cite{McG} are in  
agreement with MOND.
  
Departure of MOND from Newtonian theory is not connected with a distance 
scale. In MOND, deviations from Newtonian theory 
become significant for very small accelerations.
The main argument presented  
is that Newtonian gravitation has not been tested for very 
weak fields and it could be that the theory is not valid in this regime. 
Other theories modifying Newtonian theory were reviewed by \cite{B1} (1987).
  
Stability of galactic disks in BM theory was first studied in a WKB 
approximation by \cite{MM}. The WKB scheme deals with perturbation
wavelengths $\lambda=2\pi/k$, much smaller than the involved distances $\rho$ 
on the disk ($|k\rho|>>1$). 
In Newtonian gravitation, the presence of a dark halo is important in 
stabilizing the disk against violent bar formation \cite{OP}. Recent 
N-body simulations for the BM theory, indicate 
that disks are more stable in MOND than in Newtonian dynamics with dark halos
(\cite{BrdM} 1999). 

We obtain the conservative 
Hamiltonian that describes the motion of a particle in a previous calculated 
potential for the BM theory in Section \ref{2}. This Section also includes a 
summary of the 
stability theory for phase-space orbits. Orbits 
in a dwarf galaxy are discussed in Section \ref{3}. 
Our conclusions are presented in Section \ref{4}.
\section{Theory \label{2}} 
 
In the BM theory, the Poisson equation for determining the gravitational potential is 
modified to  
\begin{equation} 
\nabla .\left( \mu \left( \frac{\mid \nabla\varphi \mid }{a_0}\right) 
\nabla\varphi \right) =4\pi G\rho ,  \label{mond} 
\end{equation} 
where $a_0=2\times 10^{-8}\mbox{cm}/\mbox{s}^2$ is a constant with the dimension of 
acceleration, set so as to agree with the Tully-Fisher law (Milgrom \cite{M2}), the 
function  $\mu(x)$ (where $x=|\nabla\varphi|/a_0$) obeys $0<\mu (x)<1,$ with 
$\lim_{x\rightarrow 0}\mu(x)=x$ and $G$ is the Newtonian gravitational constant.
With this particular   
value of $a_0$, non-Newtonian effects due to the Solar gravitational field 
are expected to begin only beyond the Oort cloud \cite{Oo}.

Let us assume the existence of a constant external gravitational acceleration 
$\nabla\phi_g$ due to some source (e.g., a neighboring massive galaxy, such as 
ours) and a free falling sphere of mass $M_d$ 
(e.g., the center of a dwarf galaxy). Sufficiently close to $M_d$ so that
$\mu(x)\approx 1$, the solution of equation (\ref{mond}) yields the well known 
Newtonian potential \[\varphi\approx \frac{GM_d}{r}.\] 

\placefigure{esq}

When the solution 
$\varphi$ of equation (\ref{mond}) implies gravitational accelerations much bigger than the external gravitational 
acceleration $(|\nabla\varphi|>>|\nabla\phi_g|)$ but  
$|\nabla\varphi|< a_0$,  
spherical symmetry can be assumed and the potential is obtained from equation 
(\ref{mond}) in the limit $\mu(x)\approx x$, using Gauss's theorem  
\begin{equation}
\varphi\approx \sqrt{GM_da_0}\log r .\label{sesf}
\end{equation}
In the opposite limit, when $|\nabla\varphi|<<|\nabla\phi_g|$, \cite{BM} (1984)
have calculated the approximate potential of equation (\ref{mond}) in 
cylindrical coordinates 
\begin{equation}
\varphi=-\frac{(\mu _g\sqrt{1+L_g})^{-1}M_dG}{\sqrt{z^2(1-\alpha _g)+\rho ^2}}, 
\label{pot} 
\end{equation} 
where $z$ is the direction of $\nabla\phi_g$, $\mu_g=\mu(|\nabla\phi_g|/a_0)$, 
$L_g=d\ln (\mu )/d\ln (x)\mid _{x=\mid \nabla \phi _g\mid /a_0}$ and 
$\alpha _g=L_g/(1+L_g)$. In the  
asymptotic Newtonian limit $|\nabla\phi_g|>>a_0,$ 
$\mu_g\rightarrow1$, $L_g \rightarrow0$, $\alpha_g\rightarrow0$. In the  
non-Newtonian limit, which we call the MOND limit,  
$|\nabla\phi_g|<<a_0$, 
\begin{eqnarray} 
\mu_g(x)&\rightarrow&x\label{ml1}\\ 
L_g(x)&\rightarrow& 1\label{ml2}\\ 
\alpha_g&\rightarrow& 1/2\label{ml3}. 
\end{eqnarray} 
It is to be noted that equation (\ref{pot}) is not a consequence  
of tidal effects, since the external field is constant (\cite{BM} 1984). 

We study the orbit of a very small mass $m<<M_d$ such that it does not 
modify the potential in equation (\ref{pot}) and consider the motion of $m$, 
bounded by this potential.  
This is a central force type problem, so that conservation of linear  
momentum implies that the coordinate dependence of the Hamiltonian is in the  
relative distance between $M_d$ and $m.$ According to equation (\ref{pot}), 
and  
considering the azimuthal symmetry of the potential, the dynamics of the $M_d-m$ 
system are governed by the conservative Hamiltonian 
\begin{eqnarray} 
H&=&\frac{1}{2\mu_r}\left(p_\rho^2+p_z^2+\frac{l_z^2}{\rho^2} 
\right)-\frac{\lambda}{\sqrt{z^2(1-\alpha _g)+\rho^2}}\label{H},\\ 
\lambda&=&(\mu _g\sqrt{1+L_g})^{-1}GM_dm,\label{lbd} 
\end{eqnarray} 
where $\mu_r=M_dm/(M_d+m)\approx m$ is the reduced mass and $l_z$ is the angular  
momentum component in the direction of $\nabla\phi_g$ (the direction joining 
the neighboring massive galaxy and the dwarf galaxy). 
Thus, the
three dimensional motion in a axisymmetric potential is reduced to the
motion in a plane. In equations (\ref{H}) and (\ref{lbd}), cartesian 
coordinates $(\rho,z)$ are used to describe this (non-uniformly) rotating 
plane, which is often called the meridional plane.
 
In the MOND limit, when $|\nabla\phi_g|<<a_0$, equation (\ref{pot}) with 
the condition $|\nabla\varphi|<<|\nabla\phi_g|$, yields a minimum 
distance between $M_d$ and $m$:
\begin{equation} 
r_{min}=\sqrt{\frac{GM_d}{\sqrt{2}a_0}}\frac{a_0}{|\nabla\phi_g|}\label{cdv}, 
\end{equation} 
where the values for $L_g$ and $\mu_g$ are obtained from equations (\ref{ml1})- 
(\ref{ml3}). We note that the forces in the $\rho$ and $z$ directions are  
not equal. 
 
The Hamiltonian (\ref{H}) has an elliptic equilibrium position at $z=0$,  
$\rho=l_z^2/(\mu_r\lambda)$, which is called the guiding center.  
Expanding the Hamiltonian in a power series about the guiding center, we have  
\begin{eqnarray} 
& &H=H_0(I)+H_1(I,\theta)\nonumber\\ 
& &H_0(I)=\omega_zI_z+\omega_\rho I_\rho\label{H0}\\ 
& &\omega_\rho=\frac{\lambda^2\mu_r}{l_z^3},\;\;\;\omega _z= 
\frac{\lambda^2\mu_r\sqrt{1-\alpha _g}}{l_z^3}. 
\label{freq} 
\end{eqnarray} 
This expansion is called the epicycle approximation and is appropriate 
in the neighborhood of the equilibrium. $H_0$ is the unperturbed 
motion and describes the epicycles
around the guiding center; the two frequencies $\omega_\rho$ and
$\omega_z$ given in equation (\ref{freq}), are called the epicycle frequency and
the vertical frequency, respectively.    
In phase space, the conservation integrals define a $2$-dimensional torus. 
When $j_z\omega_z+j_\rho\omega_\rho=0,\;j=|j_z|+|j_\rho|$  
for integers $j_{z,\rho}$, the linearized frequencies are said to satisfy a 
resonance of order $j$. The corresponding torus is called a resonant torus. 
From equations (\ref{ml3}) and (\ref{freq}), the allowed linearized resonances are 
restricted to $\alpha_g<1/2$. 
 
\placetable{tb1} 
 
For a resonance of order  
$j$, it is possible to put the Hamiltonian into a Birkhoff normal form of  
degree $j$ by a canonical transformation 
\[H=K_0(J)+K_1(J,\Phi),\]   
where $K_0$ is a polynomial of degree $j$ in the new actions $J$ and 
$\Phi$ are the new angles \cite{MCA}. $K_0$ is a good approximation when 
the perturbation $K_1$ is sufficiently small. The new frequencies are, then, 
\begin{equation} 
\nu_i=\frac{\partial K_0(J)}{\partial J_i} 
\end{equation} 
and are functions of the new amplitudes $J_i$, a phenomenon first  
discovered by \cite{LI} in connection with nonlinear oscillators.   
 
Nondegenerescence is defined as the nonvanishing of the   
Hessian determinant 
\begin{equation} 
\det\left|\frac{\partial\nu_i}{\partial J_k}\right|\neq 0\label{nd}. 
\end{equation}   
 
There exists a theorem of \cite{K} on the behavior of a 
nonresonant torus under a small perturbation $K_1$ of a nondegenerate  
Hamiltonian $K_0$ \cite{MCA}, which was subsequently proved by   
\cite{A1} for Hamiltonian systems and by \cite{M} for area-preserving  
maps (see also \cite{A2} and \cite{Mo}).  
The theorem, known as KAM in recognition of their work, states the existence of  
invariant tori, densely filled with phase space curves winding around them, which are   
conditionally-periodic with the number of independent frequencies equal to the  
number of degrees of freedom.  
This theorem is valid if the following condition holds
\begin{equation} 
|(\nu.j)|>C|j|^\tau\label{nr}. 
\end{equation}
In equation (\ref{nr}) $C$ is dependent on the magnitude of the 
perturbation and $\tau$, on the number of degrees of freedom.
The existence of an invariant, or  
KAM, torus surrounding a periodic orbit, characterizes the stability.  
Orbits whose frequencies satisfy equation (\ref{nr}), occupy most of the volume 
in phase space for sufficiently small $C$ (\cite{Mo}).  

The conserved $l_z$ is an isolating integral for the full three degrees of 
freedom system;
the two degrees of freedom system described by equation (\ref{H}), is a subspace of 
constant $l_z$ of the original three degrees of freedom system.
A surface of section can be constructed in the spirit of \cite{HH} (1964).  
In a two degrees of freedom system, for instance equation (\ref{H}), the dimension of  
the phase  
space is four. Since the energy is an isolating integral, motion is 
constrained to a  
three dimensional constant energy surface, for example, $\rho$, $z$ and  
$p_z$. Successive intersections of the trajectory with  
the plane $\rho=\mbox{const.}$ and $p_\rho>0$, described by the set of points $z$ and  
$p_z$ (or $\dot{z}=dz/dt$), is called a surface of section.  
Each intersection of the orbit is a point in this plane and the passage of one  
point to the next can be considered as a mapping. After an infinite time interval, 
the points corresponding to a unique orbit fill up a whole area of the 
surface of section, in the absence of further isolating integrals.

Motion is constrained to the 
intersection of the surfaces of constant $H$ and $I$ in the presence of a 
third isolating integral. Consider an 
orbit whose initial conditions lie in the phase space region of conserved $I$. 
Its points in the surface of section form a smooth curve.
 
According to the Poincare-Birkhoff theorem for resonant tori, 
after the inclusion of the perturbation $K_1$, we have an even number of 
periodic orbits in the vicinity of a stable periodic orbit.
Periodic orbits are 
exact resonances of the nonlinear problem.  
The corresponding surface of section is a fixed point, surrounded by an 
even number of other fixed points.
Half of them are stable, while  
the other half are unstable. Stable points are surrounded by closed invariant  
curves. Unstable points are connected by separatrices, forming a  
structure called a chain of islands. 
Chaos is present in the vicinity of the separatrices.   
(For further details, see \cite{LL} and \cite{DN}.) 
  
\section{Orbits in a Dwarf Galaxy \label{3}} 
 
In the following, we describe the motion of a particle $m$ (e.g. a globular 
cluster) around a spheroid of mass $M_d$ (the nucleus of a dwarf galaxy) with 
$M_d>>m$, as shown in Figure \ref{esq}. We take    
\begin{eqnarray} 
M_d&=&10^8M_\odot,\nonumber\\ 
m&=&10^5M_\odot. \label{massas} 
\end{eqnarray} 

We assume that the $M_d-m$ system is gravitationally bounded to a nearby 
massive galaxy
$M_g=5 \times 10^{11}M_\odot$ (such as the Milky Way), separated by a distance  
$d=100kpc$. From equation (\ref{sesf}), the nearby massive galaxy  generates a 
gravitational acceleration on the order of  
\begin{equation} 
|\nabla\phi_g|\sim\sqrt{GM_ga_0}/d\sim 5.36^{-1}a_0,\label{condi1}  
\end{equation} 
within the $M_d-m$ system. The period of the test particle is appreciably less 
than that of the dwarf galaxy's orbit.
Possible resonances between the dwarf galaxy's orbital 
period and that of the test particle are neglected. According to equation 
(\ref{cdv}), 
equation (\ref{pot}) is valid for distances greater than $r_{min}=1189$pc and 
we choose  
\begin{equation} 
r\approx 2378\mbox{pc} \label{condi2}.   
\end{equation} 
Tidal accelerations for distances such as $r$ are neglected,  
so that $\nabla\phi_g$ is considered to be a constant external gravitational  
acceleration within the $M_d-m$ system. The geometry is shown in  
Figure \ref{esq}. 
   
If the initial conditions are chosen as $z=p_z=0$, according to equation (\ref{H}),  
Kepler's Theory is recovered with an effective gravitational constant 
(eq. (\ref{lbd})).  
These initial conditions correspond to orbits constrained to the equator of $M_d$  
and are known as equatorial orbits. (The equatorial plane is 
defined as the plane containing $M_d$, perpendicular to the $z$ direction 
(see Fig. \ref{esq}).) 
It then follows that
\begin{eqnarray} 
l_z^2&=&\mu_r\lambda a(1-e^2)\label{exc}\\ 
H&=&E=-\frac{\lambda}{2a},\label{sem}\\ 
T&=&2\pi a^{3/2}\sqrt{\frac{\mu_r}{\lambda}}, \label{Kpl} 
\end{eqnarray}  
where $a$ is the semi-major axis, $e$ the eccentricity, $T$ the period and $E$,  
the energy. In Newtonian theory, the relation between the period and the  
semi-major axis is Kepler's third law. The equatorial orbits are uniquely  
defined by equations (\ref{exc}) and (\ref{sem}).  
 
In the MOND limit, the 
period predicted by equation (\ref{pot}), for equatorial orbits, follows from  
equations (\ref{ml1})-(\ref{ml3}), (\ref{lbd}) and (\ref{Kpl}).    
\[
T\approx2\pi r^{3/2}\sqrt{\frac{\sqrt{2}|\nabla\phi_g|/a_0}{G(M_d+m)}}, \] 
which, according to equations (\ref{massas}), (\ref{condi1}) and 
(\ref{condi2}), results  
in 
\begin{equation}
T\approx 5.6\times 10^{8}\mbox{yr}\label{periodo}. 
\end{equation}
 
The physical configuration defined by equations (\ref{massas})  and 
(\ref{condi1}) is described in the following. We assume the MOND  
limit in equations (\ref{ml1}) and (\ref{ml2}) and choose  
$\alpha_g=7/16$, corresponding to the linearized lower order resonance 
of $3/4$ in Table \ref{tb1}. Under these conditions, the surfaces of section for  
the phase orbits of equation (\ref{H}) (Figs. \ref{f4} - \ref{fg2}) are  
obtained for an energy $E=-6.817\times 10^{50}\mbox{erg}$, which corresponds 
to distances of the order of that in equation (\ref{condi2}). As a numerical 
check, the energy was found to be conserved to one part in $10^{10}$.   
 
Regions of instability increase for smaller values of $l_z$. Equatorial orbits 
are defined as $z=p_z=0$ in equation (\ref{H}).
For equatorial orbits, $l_z$ is connected to the eccentricity by equation 
(\ref{exc}). According to equation (\ref{nr}), for equatorial orbits with 
sufficiently small eccentricities, almost the entire phase space is occupied 
by KAM tori surrounding the stable orbit, as shown in Fig. \ref{f4}. The 
surface of section in Fig. \ref{f4} is obtained for  
$l_z/\mu_r=57.09\mbox{kpc}\:\mbox{km}/\mbox{s}$, which corresponds to an 
equatorial orbit with eccentricity $e=0.4$. It can be seen in Fig. \ref{f4} 
that this particular equatorial orbit is stable. 

\placefigure{f4}

The MOND potential given in equation (\ref{pot}) is valid for distances 
greater than $r_{min}$, defined in equation (\ref{cdv}). This potential 
predicts equatorial orbits which belong 
to a plane perpendicular to the external gravitational acceleration produced by 
$M_g$ (see Fig. \ref{esq}). We  
numerically found that the equatorial orbits are stable for eccentricities 
$e<0.65$. 

With decreasing values of $l_z$, for example, 
$l_z/\mu_r=41.20\mbox{kpc}\:\mbox{km}/\mbox{s}$, the equatorial orbit becomes 
unstable. There exist other stable periodic orbits, 
which are not in the plane $z=0$ (see Fig. \ref{fg2}). 

\placefigure{fg2}

Associated with the upper and lower stable islands in Fig. \ref{fg2}, 
there is an exact $2:1$ resonance, which 
is a periodic orbit in the meridional plane, defined by 
$\rho,z$ coordinates. In three dimensional space $x,y,z$, this 
periodic orbit corresponds to a quasi-periodic orbit in a spatial 
region, which is a rounded cylindrical 
shell surrounding $M_d$, as shown in  Figure \ref{orb}. A closed orbit in three 
dimensional space requires an exact resonance between the azimuthal, vertical 
and epicycle frequencies.

\placefigure{orb}  
 
\section{Conclusions \label{4}}

There exists a previously calculated potential $\varphi$ in the MOND theory, 
given by equation (\ref{pot}), for a free falling sphere of mass $M_d$ in a 
constant external gravitational acceleration $\nabla\phi_g$. $\varphi$ is 
valid when $|\nabla\phi_g|>>|\nabla\varphi|$. We assume that the existence of a 
second mass $m<<M_d$ does not modify the potential. 

The system $M_d-m$, bounded by the potential $\varphi$, is described by the 
Hamiltonian given in equation (\ref{H}). Since the potential is 
axially symmetric, only the component of the angular momentum in the 
direction of $\nabla\phi_g$, $l_z$, is conserved. As a consequence, 
when $z=p_z=0$ and $l_z$ is the total angular momentum, motion is periodic and 
occurs in a plane perpendicular to $\nabla\phi_g$, which we call the equatorial 
plane  (Fig. \ref{esq}). Kepler's theory is then 
recovered with an effective gravitational constant (eq. (\ref{lbd})), and $l_z$ 
is related to the eccentricity by equation (\ref{exc}). 

We analyze the phase space of $H$, using the numerical technique of 
surfaces of section. It is found that for eccentricities
$\epsilon<0.65$, the equatorial orbits are stable with respect to small 
variations in the initial conditions (Fig. \ref{f4}). For smaller values of 
$l_z$, the stability of the equatorial orbit is lost and there is an 
increase in the size of the chaotic regions (Fig. \ref{fg2}). 
There is a disk perpendicular to $\nabla\phi_g$, as well as some  
regions not in the disk, which are stable against variations of initial 
conditions.

\acknowledgments 
D. M. would like to thank the Brazilian agency CAPES for support and R. O.  
would like to thank the Brazilian agency CNPq and the project 
PRONEX/FINEP (no. 41.96.0908.00) for partial support. D. M.  
would also like to thank the Dynamics Group at IAG-USP for  
enlightening conversations. 
 
\clearpage 
\begin{deluxetable}{cc} 
\footnotesize 
\tablecaption{Resonances \label{tb1}} 
\tablewidth{0pt} 
\tablehead{\colhead{$\alpha_g$} & \colhead{$\omega_z/\omega_{\rho}$}} 
\startdata 
$7/16$ & $3/4$\\ 
$9/25$ & $4/5$\\ 
$11/36$ & $5/6$\\ 
$24/49$ & $5/7$\\ 
$...$ & $...$ 
\enddata 
\end{deluxetable} 
 
\clearpage 
\figcaption[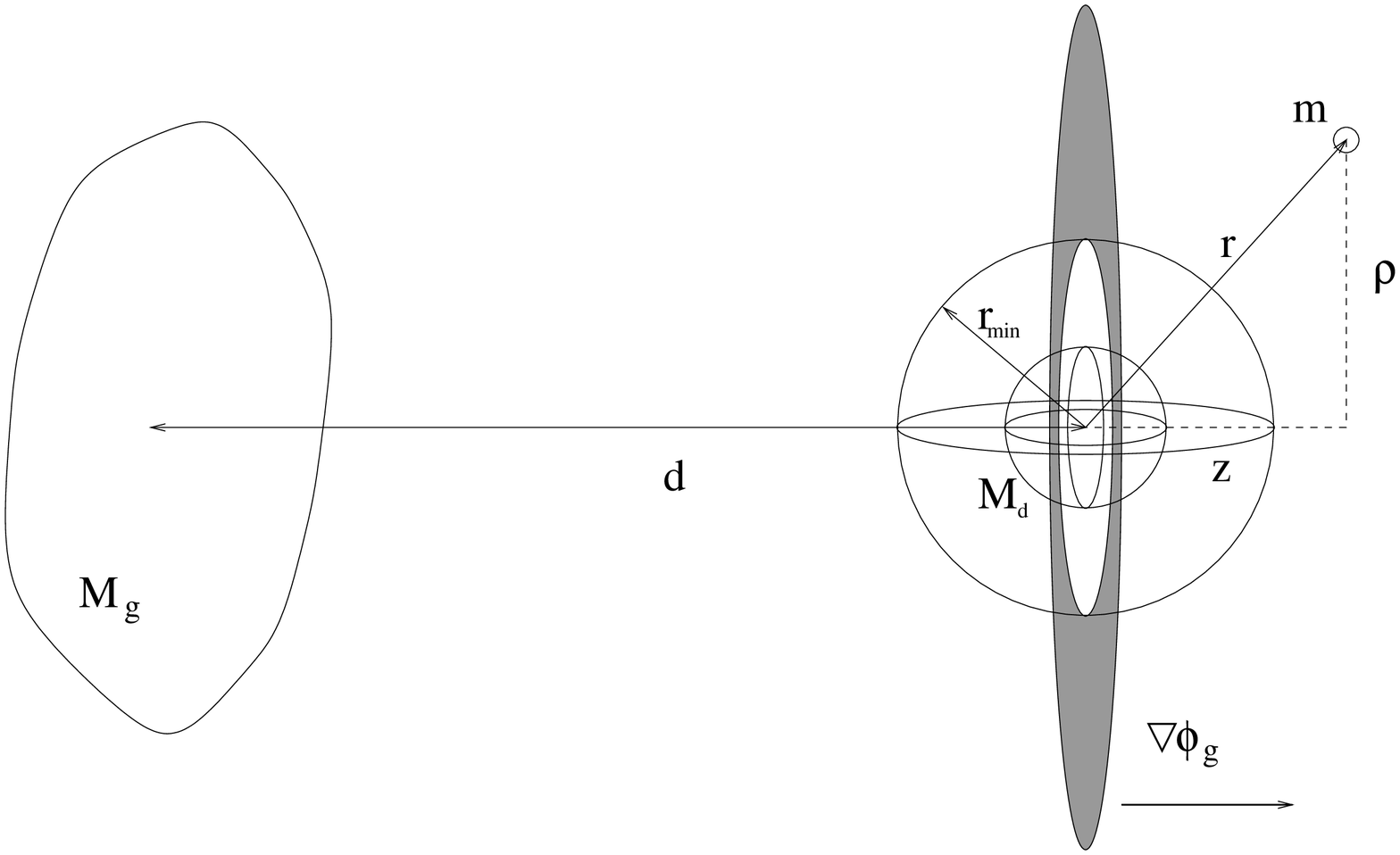]{ Schematic diagram, showing the two masses 
$M_d=10^8 M_\odot$ (e.g., nucleus of a dwarf galaxy) and $m=10^5 M_\odot$ (e.g.,  
a globular cluster) and the neighboring galaxy 
$M_g=5\times 10^{11}M_\odot$ (e.g., the Milky Way).  
According to the potential given in equation (\ref{sesf}), the 
$M_d-m$ system is gravitationally bounded to $M_g$, at 
an approximate distance of $d=100$kpc. The external gravitational acceleration 
field due to $M_g$ is $|\nabla\phi_g|\sim 5.36^{-1}a_0$, 
where $a_0=2\times 10^{-8}\mbox{cm}/\mbox{s}^2$ is the MOND constant.
The $z$ direction is the projection 
of the relative distance $r$ between $M_d$ and $m$ in the direction of the 
external field $\nabla\phi_g$ and $\rho$ is the orthogonal projection. The
equatorial plane is defined by $z=0$ and is shown in gray. $r_{min}$ is the 
minimum distance for which the potential in equation (\ref{pot}) is valid. 
\label{esq}}
\figcaption[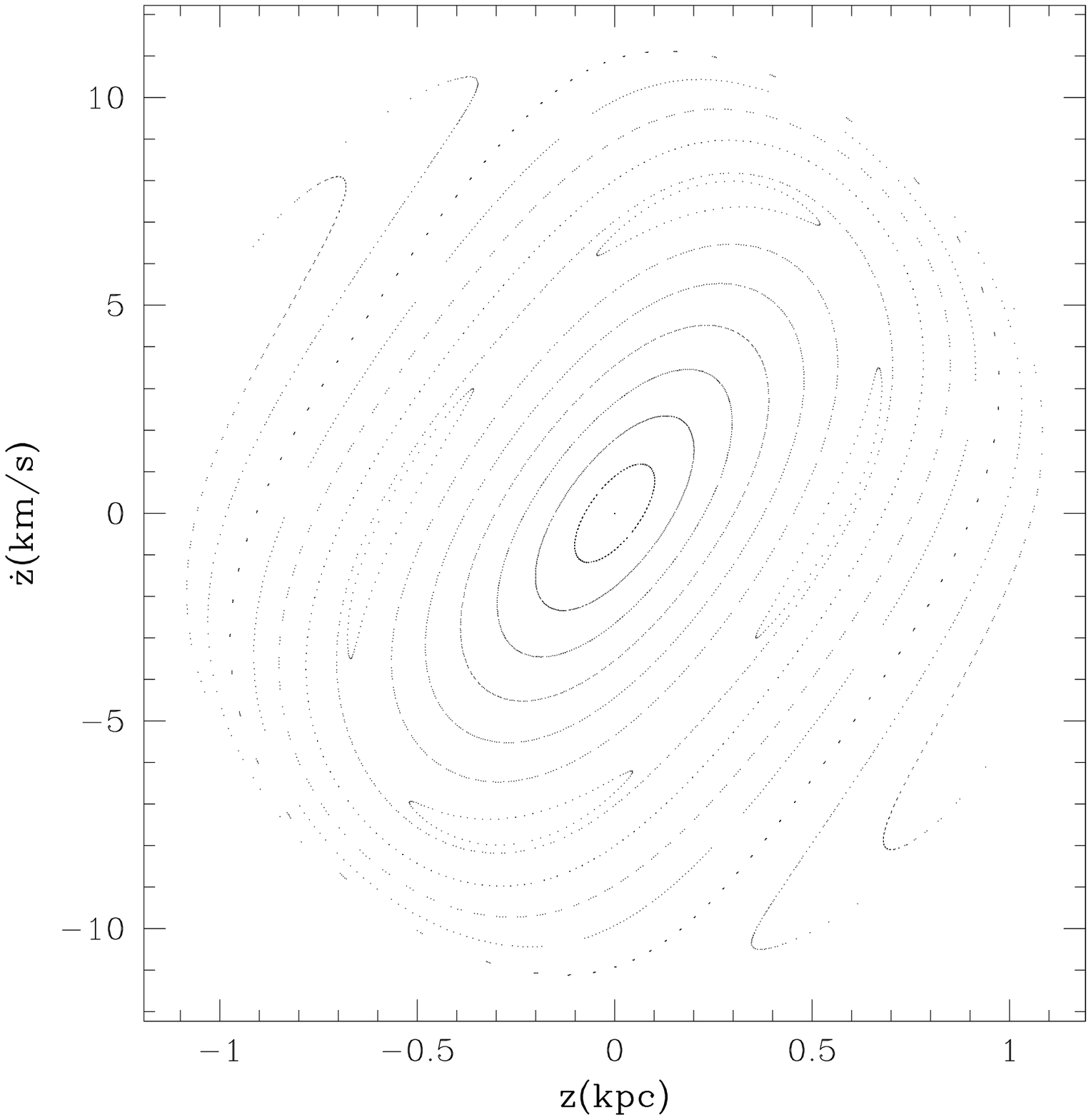]{Surface of section, for $E=-6.817\times 10^{50}\mbox{erg}$, 
and $l_z/\mu_r=57.090\mbox{kpc}\:\mbox{km}/\mbox{s}$.
\label{f4}}
\figcaption[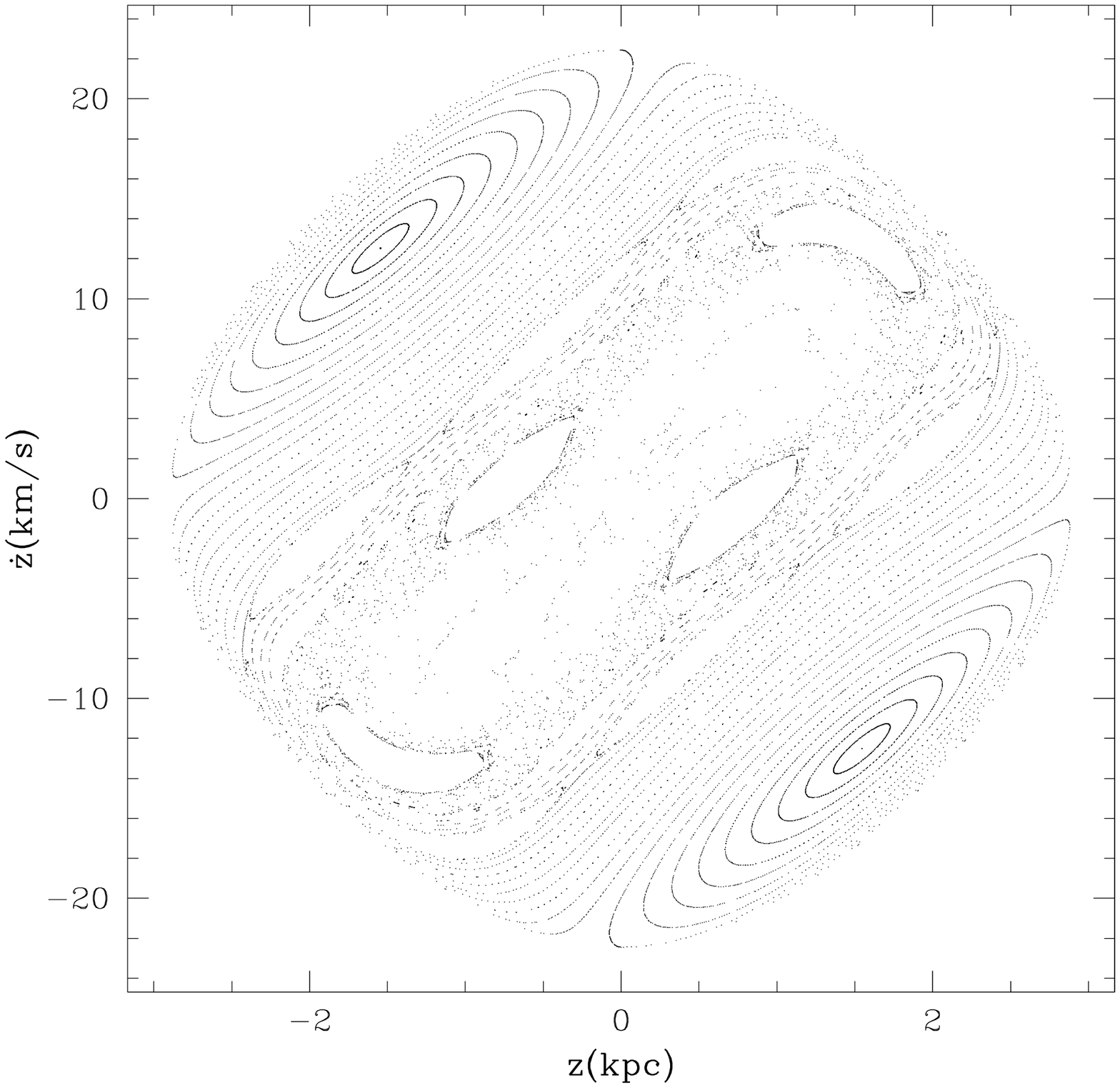]{Surface of section, for $E=-6.817\times 10^{50}\mbox{erg}$, 
and $l_z/\mu_r=41.201\mbox{kpc}\:\mbox{km}/\mbox{s}$. 
\label{fg2}} 
\figcaption[f4.ps]{ Orbit in $(x,y,z)$ space in units of kpc, corresponding to 
the $2:1$ resonance of  Figure \ref{fg2}. \label{orb}}

\clearpage 
\begin{thebibliography}{99} 
\bibitem[(Arnold 1989)]{MCA} Arnold, V.I. 1989, Mathematical  
Methods of Classical Mechanics, (New York: Springer-Verlag) 
\bibitem[Arnold (1961)]{A1} Arnold, V.I. 1961, Soviet Math. Dokl., 2, 501 
\bibitem[Arnold 1963]{A2} Arnold, V.I. 1963, Russian Math. Surv., 18:6,85 
\bibitem[Bekenstein \& Milgrom]{BM}  Bekenstein, J. D., Milgrom, M. 1984,  
\apj, 286, 7  
\bibitem[Bekenstein]{B1}  Bekenstein, J. D. 1987, in: 2nd Canadian Conference on 
General Relativity and Relativistic Astrophysics, ed. C. Dyer, (Singapore: World 
Scientific)
\bibitem[de Block \& McGaugh (1998)]{Block}de Block, W.J., McGaugh, S.S. 1998, 
\apj, 508, 132
\bibitem[Brada \& Milgrom]{BrdM} Brada, R., Milgrom, M. 1999, \apj, 519,
590
\bibitem[Dunby (1971)]{DN} Danby J. M. A. 1971,  
Qualitative Methods in Celestial Mechanics, (S\~ao Paulo: Publica\c c\~oes do Instituto 
de Matem\'atica e Estat\'\i stica da Universidade de S\~ao Paulo)  
\bibitem[Faber \& Jackson (1976)]{FJ}  Faber, S.M., Jackson, R.E. 1976, 
\apj, 204, 668 
\bibitem[H\'enon \& Heiles]{HH}  H\'enon M., Heiles C. 1964, \aj, 69, 73 
\bibitem[Kolmogorov (1954)]{K} Kolmogorov, A.N. 1954, Dokl. Akad. Nauk. SSSR,  
98, 527 
\bibitem[Lichtenberg \& Lieberman (1991)]{LL}  Lichtenberg A.J. \& 
Lieberman M.A. 1983, Regular and 
Stochastic Motion Second Edition, (New York: Springer-Verlag)  
\bibitem[Lindstedt (1882)]{LI}  Lindstedt, M. 1882, Astrom. Nach., 103, 211 
\bibitem[McGaugh \& de Block (1998)]{McG} McGaugh, S.S., de Block, W.J.G. 1998, \apj, 
499, 66
\bibitem[(Milgrom 1983a,]{M1}  Milgrom, M. 1983a, \apj, 270, 365 
\bibitem[1983b]{M2}  Milgrom, M. 1983b, \apj, 270, 371 
\bibitem[1983c)]{M3}  Milgrom, M. 1983c, \apj, 270, 384
\bibitem[Milgrom (1989)]{MM} Milgrom, M. 1989, \apj, 338, 121 
\bibitem[Moser (1962)]{M} Moser, J. 1962, On Invariants Curves of  
Area-Preserving Mappings on an Annulus, Nachr. Akad. Wiss.  
G\"otingen, Math. Phys. K1, p.1 
\bibitem[Moser 1971]{Mo} Moser, J., Siegel, C. 1971, Lectures in Celestial  
Mechanics, (Berlin: Springer-Verlag) 
\bibitem[(Oort 1963)]{Oo} Oort, J.H. 1963, The Moon, Meteorites, and  
Comets, ed. B. M. Middlehurst and G. P. Kuiper (Chicago:  
University of Chicago Press)
\bibitem[(Ostriker \& Peebles 1973)]{OP} Ostriker, J.P., Peebles, P.J.E. 1973, 
\apj, 186, 467
\bibitem[Salcedo \& G\'amez (1999)]{SG} Salcedo, F.J.S., G\'amez, A.M.H. 1999, 
\aap, 345, 36 
\bibitem[Tully \& Fisher (1977)]{TF}  Tully, R.B., Fisher, J.R. 1977, \aap, 
54, 661   
\end{thebibliography}
\end{document}